\documentclass[12pt,letterpaper]{article}
\usepackage{jheppub}

\title{New Local Duals in Eternal Inflation}

\author[a,b]{Raphael Bousso,}
\affiliation[a]{Center for Theoretical Physics and Department of Physics,\\
 University of California, Berkeley, CA 94720, U.S.A.}
\affiliation[b]{Lawrence Berkeley National Laboratory, Berkeley, CA 94720,
  U.S.A.}

\author[a,b]{and Dan Mainemer Katz}

\abstract{Global-local duality is the equivalence of seemingly different regulators in eternal inflation.  For example, the light-cone time cutoff (a global measure, which regulates time) makes the same predictions as the causal patch (a local measure that cuts off space).  We show that global-local duality is far more general.  It rests on a redundancy inherent in any global cutoff: at late times, an attractor regime is reached, characterized by the unlimited exponential self-reproduction of a certain fundamental region of spacetime.  An equivalent local cutoff can be obtained by restricting to this fundamental region.  

We derive local duals to several global cutoffs of interest.  The New Scale Factor Cutoff is dual to the {\em Short Fat Geodesic}, a geodesic of fixed infinitesimal proper width.  Vilenkin's CAH Cutoff is equivalent to the {\em Hubbletube}, whose width is proportional to the local Hubble volume.  The famous youngness problem of the Proper Time Cutoff can be readily understood by considering its local dual, the {\em Incredible Shrinking Geodesic}.}

\begin{document}
\maketitle

\section{Introduction}

Global-local duality is one of the most fascinating properties of eternal inflation.  It lies at the heart of a profound debate: should we expect to understand cosmology by adopting a global, ``bird's eye'' viewpoint that surveys many causally disconnected regions, as if we stood outside the universe as a kind of meta-observer? Or should fundamental theory describe only experiments that can be carried out locally, in accordance with the laws of physics and respectful of the limitations imposed by causality?

Global-local duality appears to reconcile these radically different perspectives.  It was discovered as a byproduct of attempts to solve the {\em measure problem} of eternal inflation: the exponential expansion of space leads to infinite self-reproduction, and all events that can happen in principle will happen infinitely many times.\footnote{The measure problem has nothing to do with how many vacua there are in the theory.  It arises if there exists at least one stable or metastable de~Sitter vacuum.  The observed accelerated expansion of the universe~\cite{Per98,Rie98} is consistent with a fixed positive cosmological constant and thus informs us that our vacuum is likely of this type~\cite{Bou12}.} (Short reviews include Refs.~\cite{Fre11,Sal11}.)  To compute relative probabilities, a regulator, or measure, is required.  

Most measure proposals are based on geometric cutoffs: one constructs a finite subset of the eternally inflating spacetime according to some rule.\footnote{As a consequence, probabilities behave as if the spacetime was extendible~\cite{BouFre10c,GutVan11}, a counterintuitive feature that underlies the phenomenological successes of some measures.} Relative probabilities can then be defined as ratios of the expected number of times the corresponding outcomes that occur in these subsets.  Geometric cutoffs can be divided into two classes.  Very roughly speaking, global cutoffs act on time, across the whole universe; this is natural from the bird's eye viewpoint.  Local cutoffs act on space; this is more natural from the viewpoint of an observer within the spacetime.  

Global cutoffs define a parameter $T$ that can roughly be thought of as a time variable.  Spacetime points with $T$ smaller than the cutoff form a finite set in which expected numbers of outcomes can be computed; then the limit $T\to\infty$ is taken to define probabilities:
\begin{equation}
\frac{P_I}{P_J}\equiv \lim_{T\to\infty}\frac{N_I(T)}{N_J(T)}~.
\end{equation}
Examples include the proper time cutoff~\cite{Lin86a,LinLin94,GarLin94,GarLin94a,GarLin95}, where $T$ is the proper time along geodesics in a congruence; the scale factor time cutoff~\cite{LinMez93,LinLin94,Lin06,DesGut08a,DesGut08b,BouFre08b,Bou12b}, where $T$ measures the local expansion of geodesics; and the light-cone time cutoff~\cite{Bou09,BouYan09,BouFre10b}, where $T$ is determined by the size of the future light-cone of an event.\footnote{In the absence of a first-principles derivation, the choice between proposals must be guided by their phenomenology.  Fortunately, different definitions of $T$ often lead to dramatically different predictions (see, e.g., \cite{Lin07,BouFre10e}).  In this paper, we do not consider phenomenology but focus on formal properties.}

Local cutoffs restrict to the neighborhood of a single timelike geodesic.  The simplest local cutoff is the causal patch~\cite{Bou06,BouFre06a}: the causal past of the geodesic, which depends only on the endpoint of the geodesic.   Another example is the fat geodesic~\cite{BouFre08b}, which restricts to an infinitesimal proper volume near the geodesic.  Relative probabilities are defined by computing an ensemble average over different possible histories of the cutoff region:
\begin{equation}
\frac{P_I}{P_J}\equiv \frac{\langle N_I \rangle}{\langle N_J \rangle}~.
\end{equation}

Global-local duality is the statement that there exist pairs of cutoffs---one global, one local---that yield precisely the same predictions.  Our goal will be to exhibit the generality of this property and the basic structure underlying it.  This will allow us to identify new local duals to some global measures of particular interest.  

\paragraph{Discussion} Global-local duality implies that the distinction between two seemingly disparate perspectives on cosmology is, at best, subtle.  However, it is too early to conclude that the global and local viewpoints are as interchangeable as the position and momentum basis in quantum mechanics.  Some important distinctions remain; and for now, each side, global and local, exhibits attractive features that the other lacks.

{\em Advantages of the global viewpoint:} A key difference between global and local cutoffs is that local measures are sensitive to initial conditions, whereas global measures exhibit an attractor regime that completely determines all probabilities.  The attractor regime can only be affected by infinite fine-tuning, or by choosing initial conditions entirely in terminal vacua so that eternal inflation cannot proceed.  Thus, global measures are relatively insensitive to initial conditions.  

Therefore, a local cutoff can reproduce the predictions of its global dual only with a particular choice of initial conditions on the local side, given by the attractor solution of the global cutoff.  (The distribution over initial conditions is set by the field distribution on a slice of constant global cutoff parameter.) Ultimately, there appears to be no reason why initial conditions might not be dictated by aspects of a fundamental theory unrelated to the measure.  In this case, global and local measures could be inequivalent.  But for now, the global cutoff is more restrictive, and thus more predictive, than its local dual, some of whose predictions could be changed by a different choice of initial conditions.\footnote{In a theory with a vacuum landscape large enough to solve the cosmological constant problem~\cite{BP}, most predictions of low-energy properties are fairly insensitive to initial conditions even in local measures~\cite{BouYan07}.}

{\em Advantages of the local viewpoint:} Through the study of black holes and the information paradox, we have learned that the global viewpoint must break down at the full quantum level.  Otherwise, the black hole would xerox arbitrary quantum states into its Hawking radiation, in contradiction with the linearity of quantum mechanics~\cite{SusTho93}.  By contrast, a description of any one causally connected region, or causal patch, will only contain one copy of the information.  For example, an observer remaining outside the black hole will be able to access the Hawking radiation but not the original copy, which is accessible to an observer inside the black hole.  Indeed, this was the key motivation for introducing the causal patch as a measure: if required in the context of black holes, surely the same restriction would apply to cosmology as well.  

The global spacetime is obtained by pretending that the state of the universe is measured, roughly once per Hubble time in every horizon volume.  It is not clear what the underlying process of decoherence is, since no natural environment is available (by definition, since we are considering the entire universe).  By contrast, the local description (at least, the causal patch) exhibits decoherence at the semiclassical level, since matter can cross the event horizon~\cite{BouSus11}.  This suggests that the global picture may be merely a convenient way of combining the different semiclassical histories of the causal patch into a single spacetime.

Aside from the fundamental questions raised by global-local duality, we expect that our results will aid future studies of measure phenomenology.   Computations are significantly simpler in the local dual, because it strips away an infinite redundancy.  The local cutoff region can be considered an elementary unit of spacetime, which (from the global viewpoint) is merely reproduced over and over by the exponential expansion of the eternally inflating universe.

\paragraph{Outline} In Sec.~\ref{sec-cp}, we set the stage by showing that the light-cone time cutoff is equivalent to the causal patch, with initial conditions in the longest-lived de~Sitter vacuum.  This is a known result~\cite{Bou09,BouYan09}. The measures on both sides of the duality are particularly simple; as a consequence, the duality proof is especially transparent.

In Sec.~\ref{sec-fg}, we define the {\em Short Fat Geodesic} measure, and we show that it is the local dual to the recently proposed New Scale Factor Cutoff~\cite{Bou12b}.  This generalizes to arbitrary eternally inflating spacetimes the known duality between the fat geodesic and the scale factor time cutoff~\cite{BouFre08b}, which originally applied only to everywhere-expanding multiverse regions.  We also discuss important formal differences between the causal patch/light-cone time pair and all other global-local pairs we consider.  Unlike other local measures, which require the specification of spatial boundary conditions, the causal patch is entirely self-contained and can be evaluated without referring to a global viewpoint.

In Sec.~\ref{sec-x}, we generalize the proof of the previous section to relate a large class of global-local pairs.  On the local side, one can consider modulations of the fatness of the geodesic; on the global side, this corresponds to particular modifications of the definition of the cutoff parameter $T$, which we identify explicitly.  We illustrate this general result by deriving local duals to two global proposals, the CAH cutoff~\cite{Vil11} and the proper time cutoff.  The local dual, the {\em Hubbletube}, naturally extends the range of applicability of the CAH cutoff to include decelerating regions; but unfortunately, an additional prescription (such as CAH+~\cite{Vil11}) is still required to deal with nonexpanding regions.  The local dual to the proper time cutoff, the {\em Incredible Shrinking Geodesic}, makes the phenomenological problems of this simplest of global cutoffs readily apparent.

\section{Causal Patch/Light-Cone Time Duality}
\label{sec-cp}

In this section we show that the causal patch measure (with particular initial conditions) is equivalent to the light-cone time measure, i.e., that both define the same relative probabilities.  We follow Ref.~\cite{BouYan09}, where more details can be found.\footnote{See Ref.~\cite{HarShe11,Sus12} for a simplified model that exhibits most of the essential features of eternal inflation, including causal patch/light-cone time duality.}  The proof is rather simple if one is willing to use the (intuitively natural) results for the attractor behavior of eternal inflation as a function of the global time coordinate.  For this reason we will first present a proof of duality, while assuming the attractor behavior, in Sec.~\ref{sec-cpdef}.  Then we will derive the attractor behavior, in Sec.~\ref{sec-lcattract}.

\subsection{Causal Patch Measure}
\label{sec-cpdef}

The causal patch is defined as the causal past of the endpoint of a geodesic.  Consider two outcomes $I$ and $J$ of a particular observation; for example, different values of the cosmological constant, or of the CMB temperature.   The relative probabilities for these two outcomes, according to the causal patch measure, is given by
\begin{equation}
\frac{\hat P_I}{\hat P_J}=\frac{\langle N_I \rangle_{\rm CP}}{\langle N_J \rangle_{\rm CP}}~,
\label{eq-cpprob}
\end{equation}
Here, $\langle N_I \rangle$ is the expected number of times the outcome $I$ occurs in the causal patch.  

Computing $\langle N_I \rangle_{\rm CP}$ involves two types of averaging: over initial conditions, $p^{(0)}_i$, and over different decoherent histories of the patch.  We can represent the corresponding ensemble of causal patches as subsets of a single spacetime.  Namely, we consider a large initial hypersurface (a moment of time), $\Sigma_0$, containing $Z\to\infty$ different event horizon regions, with a fraction $p^{(0)}_i$ of them in the vacuum $i$.  Event horizons are globally defined, but we will be interested in cases where the initial conditions have support mainly in long-lived metastable de~Sitter vacua.  Then we make a negligible error by assuming that the event horizon on the slice $\Sigma_0$ contains a single de Sitter horizon volume, of radius $H_\alpha^{-1}=(3/\Lambda_\alpha)^{1/2}$, where $\Lambda_\alpha$ is the cosmological constant of vacuum $\alpha$\footnote{We will use Greek indices to label de~Sitter vacua ($\Lambda>0$), indices $m,n,\ldots$ to label terminal vacua ($\Lambda\leq 0$), and $i,j$ for arbitrary vacua.}, and we may take the spatial geometry to be approximately flat on $\Sigma_0$.  More general initial conditions can be considered~\cite{BouYan09}.

At the center of each initial horizon patch, consider the geodesic orthogonal to $\Sigma_0$, and construct the associated causal patch.   We may define $\langle N_I \rangle_{\rm CP}$ as the average over all $Z$ causal patches thus constructed, in the limit $Z\to\infty$.  So far, each causal patch is causally disconnected from every other patch.  It is convenient to further enlarge the ensemble by increasing the density of geodesics to $z$ geodesics per event horizon volume, and to take $z\to\infty$:
\begin{equation}
\langle N_I \rangle_{\rm CP} =(zZ)^{-1} \sum_{\nu=1}^{zZ} N_I^{\nu,\rm CP} ~,
\label{eq-cpsum}
\end{equation}
where the sum runs over the $zZ$ causal patches, and $N_I^{\nu,\rm CP}$ is the number of times $I$ occurs in the causal patch $\nu$.  The causal patches will overlap, but this will not change the ensemble average.  

\begin{figure}
\begin{center}
\includegraphics[scale = .6]{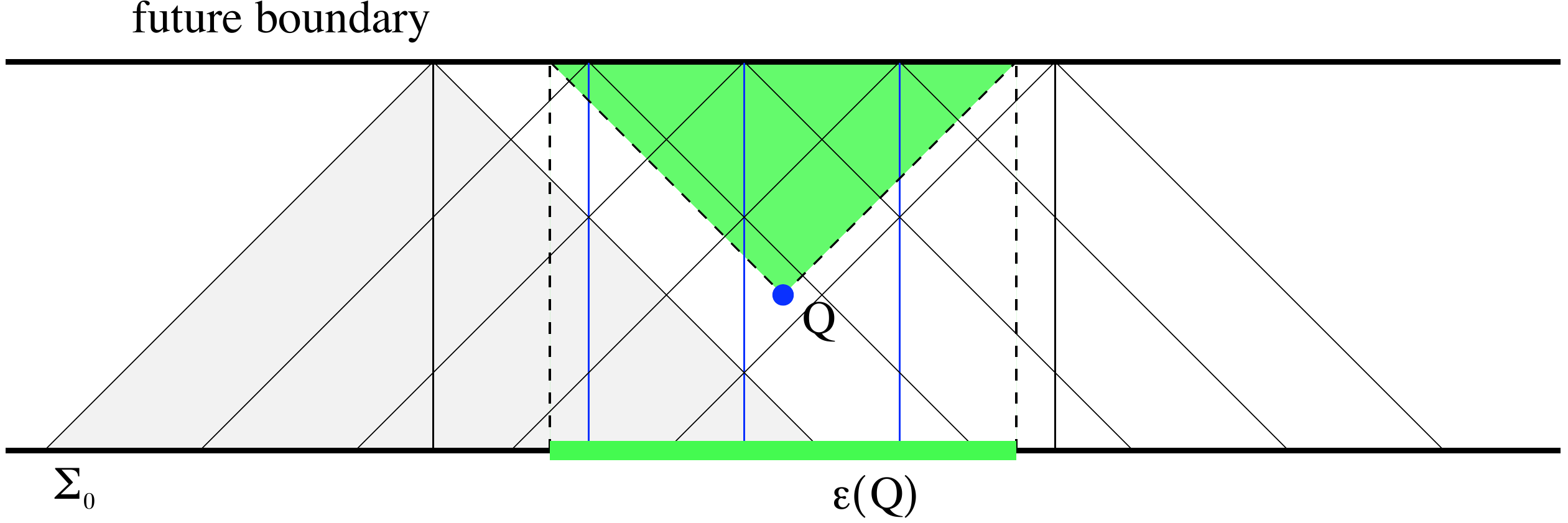}
\end{center}
\caption{Discrete ensemble of causal patches~\cite{BouYan09}.  The event $Q$ is contained in those causal patches whose
  generating geodesics (blue) enter the causal future of $Q$, $I^+(Q)$
  (shaded green/dark).  In the continuous limit, $z\to\infty$, the causal patch measure weights $Q$ in
  proportion to the volume of its future light-cone on the future boundary.  Thus, the weight of $Q$ depends only on light-cone time $t_{\rm LC}$.  This underlies the equivalence of the causal patch measure (with particular initial conditions) and the light-cone time cutoff.  This is a conformal (or Penrose) diagram; the spacetime metric is rescaled but light-rays still travel at 45 degrees.}
\label{fig-magic}
\end{figure} 
At finite large $z$, a sufficiently early event $Q$ in the future of $\Sigma_0$ will thus be contained in a number of causal patches.  The later $Q$ occurs, the fewer patches will contain it (Fig.~\ref{fig-magic}). In the limit $z\to\infty$, every event will be contained in an infinite number of patches, but there is still a sense in which later events are overcounted less.  This can be captured by defining the quantity $\pi(Q)$, as $z^{-1}$ times the number of causal patches containing a given event $Q$.   

By causality, the causal patch of a geodesic contains $Q$ if and only if that geodesic enters the future of $Q$.  Therefore, $\pi(Q)$ is the volume, measured in units of horizon volume, on $\Sigma_0$, of the starting points of those geodesics that eventually enter the future light-cone of $Q$.  This allows us to reorganize the sum in Eq.~(\ref{eq-cpsum}).  Instead of summing over causal patches, we may sum over all events $Q$ where outcome $I$ occurs, taking into account that each such instance will be ``overcounted'' by the ensemble of causal patches, by a factor proportional to $\pi(Q)$:
\begin{equation}
\langle N_I \rangle_{\rm CP} =Z^{-1} \sum_{Q\in I} \pi(Q) ~.
\label{eq-qsum}
\end{equation}

Light-cone time is defined precisely so that it is constant on hypersurfaces of constant $\pi(Q)$.  The exact definition is not essential but it is convenient to choose
\begin{equation}
t_{\rm LC}(Q)  \equiv -\frac{1}{3}\log \pi(Q)~.
\label{eq-lcdef}
\end{equation}
This defines a time variable at every event $Q$ in the future of the initial hypersurface $\Sigma_0$.  
We may reorganize the sum once more, as an integral over light-cone time:
\begin{equation}
\langle N_I \rangle_{\rm CP} =\int dt_{\rm LC} \frac{dN_I}{d t_{\rm LC}} e^{-3 t_{\rm LC}}~,
\label{eq-tsum}
\end{equation}
where $dN_I$ is the number of events of type $I$ that occur in the time interval $(t_{\rm LC} , t_{\rm LC}+d t_{\rm LC})$, and the integral ranges over the future of $\Sigma_0$.  

\subsection{Proof of Equivalence to the Light-Cone Time Measure}

So far, we have been dealing with a local measure, the causal patch.  We have merely represented the causal patch ensemble in terms of a single global spacetime.  Moreover, we have rewritten the ensemble average, as an integral over a time variable $t_{\rm LC}$, adapted to the factor $\pi(Q)$ by which events in the global spacetime are weighted in the ensemble.  

We will now show that with a particular, simple choice of initial conditions, the causal patch probabilities $\hat P_I$ (i.e., the ensemble averages $\langle N_I \rangle_{\rm CP}$) agree with the probabilities computed from a global measure, the light-cone time cutoff.  These probabilities are defined by
\begin{equation}
\frac{\check P_I}{\check P_J}=\lim_{t_{\rm LC} \to\infty} \frac{N_I(t_{\rm LC} )}{N_J(t_{\rm LC} )}~,
\label{eq-lcprob}
\end{equation}
where $N_I(t_{\rm LC})$ is the number of events of type $I$ prior to the light-cone time $t_{\rm LC}$.\footnote{Interestingly, the light-cone time cutoff was not discovered as the global dual to the causal patch.  It was proposed independently~\cite{Bou09} as a covariant implementation~\cite{BouRan01,BouLei12} of a suggestion by Garriga and Vilenkin~\cite{GarVil08} that an analogue of the UV/IR relation~\cite{SusWit98} of gauge/gravity duality~\cite{Mal97} would yield a preferred global time variable in eternal inflation.  An apparent relation to the causal patch was immediately noted~\cite{Bou09}, but the exact duality was recognized only later~\cite{BouYan09}.}

As we shall review below, the cosmological dynamics, as a function of light-cone time, leads to an attractor regime:
\begin{equation}
N_I(t_{\rm LC})=\check N_I e^{\gamma t_{\rm LC} } + O(e^{\varphi t_{\rm LC} })~,
\label{eq-lcattract}
\end{equation}
where $\varphi<\gamma<3$.  Therefore, the light-cone time probabilities are given by
\begin{equation}
\frac{\check P_I}{\check P_J}=\frac{\check N_I}{\check N_J}~.
\label{eq-lcfinal}
\end{equation}

The causal patch probabilities can also be evaluated using Eq.~(\ref{eq-lcattract}), if we choose initial conditions in the attractor regime, i.e., if we take $\Sigma_0$ to be a slice of constant, very late light-cone time.  Substituting into Eq.~(\ref{eq-tsum}), one finds
\begin{equation}
\langle N_I \rangle_{\rm CP} =\check N_I \int dt_{\rm LC} \gamma e^{(\gamma-3) t_{\rm LC}}~,
\label{eq-tsum2}
\end{equation}
Since $\gamma<3$, the integral converges to an $I$-independent constant, so relative probabilities in the causal patch measure are given by
\begin{equation}
\frac{\hat P_I}{\hat P_J}=\frac{\check N_I}{\check N_J}~.
\end{equation}
This agrees with the light-cone time probabilities, Eq.~(\ref{eq-lcfinal}).  Therefore, the two measures are equivalent.

\subsection{Light-Cone Time Rate Equation and Attractor Solution}
\label{sec-lcattract}

We will now complete the proof by deriving the attractor regime, Eq.~(\ref{eq-lcattract}).  (We will follow~\cite{BouYan09} and will make use of certain general properties of rate equations in eternal inflation~\cite{GarSch05}.) It is convenient to do this in two steps.  Treating each long-lived metastable de~Sitter vacuum as pure, empty de~Sitter space, one derives the number $n_\alpha(t_{\rm LC})$ of horizon patches of vacuum $\alpha$.  Because of the slow decays, most regions are indeed empty, and slices of constant light-cone time are spatially flat on the horizon scale.  Thus, a horizon patch at constant time $t_{\rm LC}$ can be defined as a physical volume
\begin{equation}
v_\alpha=\frac{4\pi}{3}\tau_{\Lambda,\alpha}^3~,
\end{equation}
where
\begin{equation}
\tau_{\Lambda,\alpha} \equiv \sqrt{\frac{3}{\Lambda_\alpha}}
\end{equation}
is the time and distance scale associated with the cosmological constant in vacuum $\alpha$.
The number $n_\alpha$ of horizon patches at the time $t_{\rm LC}$ is related to the physical volume $V_\alpha$ occupied by vacuum $\alpha$, as
\begin{equation}
n_\alpha(t_{\rm LC})=\frac{V_\alpha(t_{\rm LC})}{v_\alpha}~.
\end{equation}
In the second step, one focusses on the decay events in this distribution, i.e., the production of new bubbles. These bubbles can then be considered in detail.  In general they will be not be empty, and they need not have positive cosmological constant. 

The rate equation for the number of horizon patches of metastable de~Sitter vacua is
\begin{equation}
\frac{dn_\alpha}{d t_{\rm LC}} = (3-\kappa_\alpha) n_\alpha + \sum_\beta \kappa_{\alpha \beta} n_\beta~,
\label{eq-nrate}
\end{equation}
where $\kappa_{i \beta}=v_i \tau_{\Lambda,\beta} \Gamma_{i \beta}$ is the dimensionless decay rate from $\beta$ to $i$.  That is, $\Gamma_{i \beta}$ is the rate at which $i$-bubbles are produced inside the $\beta$-vacuum, per unit four-volume; and $\kappa_{i \beta}$ is the decay rate per unit horizon volume and unit de~Sitter time scale.  Also, $\kappa_\alpha\equiv \sum_i \kappa_{i\alpha}$ is the total dimensionless decay rate of vacuum $\alpha$.  We will now explain the origin of each term on the right-hand side.

The first term, $3 n_\alpha$, arises from the exponential volume growth of de~Sitter space.  In regions occupied by vacuum $\alpha$, the metric behaves locally as $ds^2=-d\tau^2+e^{2t/\tau_{\Lambda,\alpha}} d {\mathbf x}^2$,
where $t$ is proper time.  The relation between proper time and light-cone time is
\begin{equation}
d t_{\rm LC} =\frac{d\tau}{\tau_{\Lambda,\alpha}}
\label{eq-dtlcdt}
\end{equation}
in pure de~Sitter space.  In metastable de~Sitter space this relation is modified, on average, by a relative correction not exceeding $\kappa_\alpha$, which can be neglected for the purposes of the rate equation.

The second term, $-\kappa_\alpha n_\alpha$ is an effective term that takes into account the decay of vacuum $\alpha$ into other vacua.  Decays of this type proceed by the formation of a bubble of the new vacuum~\cite{CDL}.  Typically, the spherical domain wall separating the vacua will be small initially, compared to the size of the event horizon of the parent vacuum.   The domain wall will then expand at a fixed acceleration, asymptotically approaching the future light-cone of the nucleation event.  A detailed treatment of this dynamics would enormously complicate the rate equation, but fortunately an exquisite approximation is available.  Even at late times, because of de~Sitter event horizons, only a portion the of parent vacuum is destroyed by the bubble.  This portion is the causal future of the nucleation point, and at late times it agrees with the comoving future of a single horizon volume centered on the nucleation point, at the nucleation time.   Because the bubble reaches its asymptotic comoving size very quickly (exponentially in light-cone time), only a very small error, of order $\kappa_\alpha$, is introduced if we remove this comoving future, rather than the causal future, from the parent vacuum.  That is, for every decay event in vacuum $\alpha$, the number of horizon patches of type $\alpha$ is reduced by $1$ in the rate equation.  This is called the {\em square bubble approximation}. The expected number of such events is $-\kappa_\alpha n_\alpha d t_{\rm LC}$.

The third term, $\sum_\beta \kappa_{\alpha \beta} n_\beta$, captures the production of bubbles of vacuum $\alpha$ by the decay of other vacua.  The prefactor of this term is fixed by the continuity of light-cone time.  This is the requirement that the future light-cone of an event $Q_{-\epsilon}$ just prior to the nucleation event has the same asymptotic size $\pi(Q_{-\epsilon})$ as the future light-cone of an event $Q_{+\epsilon}$ just after nucleation, $\pi(Q_{+\epsilon})$, as $\epsilon\to 0$.  In the square bubble approximation, this implies that each nucleation event effectively contributes a comoving volume of new vacuum equivalent to one horizon patch at the time of nucleation.  For the reasons described in the previous paragraph, one patch has the correct comoving size to eventually fill the future light-cone of $Q_{+\epsilon}$.\footnote{Note that this implies that the physical volume removed from vacuum $\beta$ is not equal to the physical volume added to vacuum $\alpha$ by the decay. This discontinuity is an artifact of the square bubble approximation and has no deeper significance.  In the exact spacetime, the evolution of volumes is continuous.}  Thus, for every decay event in which a bubble of vacuum $\alpha$ is produced, the number of horizon patches of type $\alpha$ is increased by $1$ in the rate equation.  The expected number of such events is $\sum_\beta \kappa_{\alpha \beta} n_\beta d t_{\rm LC} $.

The rate equation (\ref{eq-nrate}) has the solution~\cite{GarSch05}
\begin{equation}
n_\alpha(t_{\rm LC})=\check n_\alpha e^{\gamma t_{\rm LC}} + O(e^{\varphi t_{\rm LC}})~,
\label{eq-nsolution}
\end{equation}
where $\varphi<\gamma< 3$.  (The case $\gamma = 3$ arises if and only if the landscape contains no terminal vacua, i.e., vacua with nonpositive cosmological constant, and will not be considered in this paper.)  Here, $\gamma\equiv 3-q$ is the largest eigenvalue of the matrix $M_{\alpha \beta}$ defined by rewriting Eq.~(\ref{eq-nrate}) as $\frac{dn_\alpha}{dt_{\rm LC}}=\sum_\beta M_{\alpha \beta} n_\beta$; and $\check n_\alpha$ is the corresponding eigenvector.  The terms of order $e^{\varphi t}$ are subleading and become negligible in the limit as $t_{\rm LC} \to \infty$.  To a very good approximation (better than $q\ll 1$), the eigenvector is dominated by the longest-lived metastable de~Sitter vacuum in the theory, which will be denoted by $*$:
\begin{equation}
\check n_\alpha\approx \delta_{\alpha*}~,
\end{equation}
and
\begin{equation}
q\approx \kappa_*
\end{equation}
is its total dimensionless decay rate.

Next, we compute number of events of type $I$ prior to the time $t_{\rm LC}$.  We assume that the events unfolding in a new bubble of vacuum $i$ depend only on $i$, but on the time of nucleation.  This is true as long as the parent vacuum is long-lived, so that most decays occur in empty de~Sitter space.  For notational convenience, we will also assume that evolution inside a new bubble is independent of the parent vacuum; however, this could easily be included in the analysis.  Then the number of events of type $I$ inside a bubble of type $i$, $dN_I/dN_i$ will depend only on the light-cone time since bubble nucleation, $u_{\rm LC}\equiv t_{\rm LC} -t_{\rm LC}^{\rm nuc}$.  Therefore, we can write
\begin{equation}
N_I(t_{\rm LC})=\kappa_{I*} n_*(t_{\rm LC})+\sum_{i\neq *}\int_0^{t_{\rm LC}}\left(\frac{dN_I}{dN_i}\right)_{t_{\rm LC}-t_{\rm LC}^{\rm nuc}} \left(\frac{dN_i}{d t_{\rm LC} }\right)_{t_{\rm LC}^{\rm nuc}} d t_{\rm LC}^{\rm nuc} ~,
\label{eq-nilct1}
\end{equation}
Because the dominant vacuum $*$ plays a role analogous to an equilibrium configuration, it is convenient to separate it out from the sum, and to define $\kappa_{I*}$ as the dimensionless rate at which events of type $I$ are produced in $*$ regions.  The rate at which vacua of type $i$ are produced is
\begin{equation}
\frac{dN_i}{d t_{\rm LC}} = \sum_\beta \kappa_{i\beta} n_\beta~.
\end{equation}
By changing the integration variable to $u_{\rm LC}$ in Eq.~(\ref{eq-nilct1}), and using Eq.~(\ref{eq-nsolution}), one finds that 
\begin{equation}
N_I(t_{\rm LC})=\left(\kappa_{I*} \check n_*+\sum_{i\neq *}\sum_\beta N_{Ii} \kappa_{i\beta} \check n_\beta\right) e^{\gamma t_{\rm LC}}+O(e^{\varphi t_{\rm LC}})~,
\label{eq-NIsolution}
\end{equation}
where
\begin{equation}
N_{Ii}\equiv \int_0^\infty du_{\rm LC} e^{-\gamma u_{\rm LC}} \left( \frac{dN_I}{dN_i}\right)_{u_{\rm LC}}
\label{eq-nii}
\end{equation}
depends only on $I$ and $i$.  The above integral runs over the interior of one $i$-bubble, excluding regions where $i$ has decayed into some other vacuum.  Naively, the integral should range from $0$ to $t_{\rm LC}$.  But the global measure requires us to take the limit $t_{\rm LC}\to\infty$ in any case, and it can be done at this step separately without introducing divergences.  Since $*$ does not appear in the sum in Eq.~(\ref{eq-NIsolution}), and all other vacua decay faster than $*$, the interior of the $i$-bubble in Eq.~(\ref{eq-nii}) grows more slowly than $e^{\gamma t_{\rm LC}}$.  Therefore, the integral converges, and we may write
\begin{equation}
N_I(t_{\rm LC} )=\check N_I e^{\gamma t_{\rm LC}}+O(e^{\varphi t_{\rm LC}})~,
\end{equation}
where
\begin{equation}
\check N_I\equiv \kappa_{I*} \check n_*+\sum_{i\neq *}\sum_\beta N_{Ii} \kappa_{i\beta} \check n_\beta~.
\label{eq-checkni}
\end{equation}

\section{Short Fat Geodesic/New Scale Factor Cutoff Duality}
\label{sec-fg}

In this section we introduce the Short Fat Geodesic measure.  We show that, with particular initial conditions, it is equivalent to the New Scale Factor Cutoff~\cite{Bou12b}.  This generalizes to arbitrary eternally inflating spacetimes the duality between the (long) fat geodesic and (old) scale factor time cutoff discovered in Ref.~\cite{BouFre08b}, which applied only to everywhere-expanding multiverse regions.

\subsection{Short Fat Geodesic Measure}
\label{sec-fgdef}

A fat geodesic is defined as an infinitesimal neighborhood of a geodesic.  At each point on the geodesic, one can define an orthogonal cross-sectional volume $dV$, which we imagine to be spherical.  It is important to note that an orthogonal cross-section can be defined {\em only} infinitesimally---there is no covariant way of extending the cross-section to a finite volume.  For example, the spacelike geodesics orthogonal to a point on the geodesic in question need not form a well-defined hypersurface.

Consider a family of geodesics orthogonal to an initial hypersurface $\Sigma_0$. Along each geodesic, we may define the {\em scale factor parameter}
\begin{equation}
\eta\equiv \int \frac{\theta(\tau)}{3} d\tau~,
\label{eq-sft}
\end{equation}
where
\begin{equation}
\theta\equiv \frac{d}{d\tau}\log\frac{dV}{dV_0}
\label{eq-exp}
\end{equation}
is the expansion of the congruence.  In Eq.~(\ref{eq-exp}), $dV$ is the volume element at the proper time $\tau$ along a geodesic spanned by infinitesimally neighboring geodesics in the congruence; $dV_0$ is the volume element spanned by the same neighbors at $\tau=0$.  In terms of the unit tangent vector field (the four-velocity) of the geodesic congruence, $\xi=\partial_\tau$, the expansion can be computed as~\cite{Wald}
\begin{equation}
\theta=\nabla_a\xi^a~.
\end{equation}
If geodesics are terminated at the first conjugate point\footnote{also called focal point, or caustic; this is when infinitesimally neighboring geodesics intersect}, this procedure assigns a unique scale factor parameter to every event in the future of $\Sigma_0$~\cite{Bou12b}

A {\em Short Fat Geodesic\/} is a fat geodesic restricted to values of the scale factor parameter larger than that at $\Sigma_0$, which we may choose to be zero.   Thus, it consists of the portions of the fat geodesic along which neighboring geodesics are farther away than they are on $\Sigma_0$.  Typically, the congruence will expand locally for some time.  Eventually, all but a set of measure zero of geodesics will enter a collapsing region, such as a structure forming region such as ours, or a crunching $\Lambda<0$ vacuum.  In such regions, focal points will be approached or reached, where $\eta\to -\infty$.  The Short Fat Geodesic is terminated earlier, when $\eta=0$.\footnote{For simplicity, we will assume that the congruence does not bounce, i.e., first decrease to negative values of $\eta$ and then expand again without first reaching a caustic.  This would be guaranteed by the strong energy condition, but this condition is not satisfied in regions with positive cosmological constant.  However, it is expected to hold in practice, since the cosmological constant cannot counteract focussing on sufficiently short distance scales.} 

If the congruence is everywhere expanding, the Short Fat Geodesic reduces to the (long) fat geodesic defined in Ref.~\cite{BouFre08b}, as a special case.  This is precisely the case in which the old scale factor time is well-defined and a duality between (long) fat geodesic and old scale factor time cutoff was derived.  The duality derived below is more general and applies to arbitrary eternally inflating universes.  If the expanding phase is sufficiently long, the terminal point where $\eta=0$ can be less than one Planck time from the caustic~\cite{Bou12b}.  This is expected to be generic if the initial conditions are dominated by a long-lived metastable de~Sitter vacuum.  In this approximation, the short fat geodesic could be defined equivalently as being terminated at the first caustic.

Let us pause to point out some important differences between the causal patch cutoff discussed in the previous section, and the Short Fat Geodesic.
\begin{itemize} 

\item The causal patch depends only on the endpoint of the geodesic. It has (and needs) no preferred time foliation.  That is, there is no preferred way to associate to every point along the generating geodesic a particular time slice of the causal patch containing that point. By contrast, a specific infinitesimal neighborhood is associated to every point on the Short Fat Geodesic, so the contents of the cutoff region depend on the entire geodesic.  (The same will be true for the $X$-fat Geodesic considered in the following section.) 

\item As a consequence, the geodesic congruence could be eliminated entirely in the construction of the causal patch ensemble, in favor of a suitable ensemble of points on the future conformal boundary of the spacetime~\cite{BouFre10b}.  By contrast, the congruence is an inevitable element in the construction of all other measures considered in this paper.

\item The causal patch can be considered on its own, whereas the Short Fat Geodesic is naturally part of a larger spacetime.  In the construction of an ensemble of causal patches in Sec.~\ref{sec-cpdef}, the global viewpoint was optional.  This is because the causal patch is self-contained: if the initial state is a long-lived de~Sitter vacuum, no further boundary conditions are required in order to construct the decoherent histories of the patch.  We chose a global representation (a large initial surface with many horizon patches) only with a view to proving global-local duality.  By contrast, the Short Fat Geodesic is greater than the domain of dependence of its initial cross-section.  It has timelike boundaries where boundary conditions must be specified.  The simplest way to obtain suitable boundary conditions is from a global representation in terms of geodesics orthogonal to some surface $\Sigma_0$.  

\end{itemize}

We will consider a dense family (a congruence) from the start, because of the infinitesimal size of the fat geodesic.  We index each geodesic by the point $x_0\in\Sigma_0$ from which it originates.   For the same reason, it will be convenient to work with a (formally continuous) distribution $D_I(x)$ of events of type $I$. The distribution is defined so that the number of events of type $I$ in a spacetime four-volume $V_4$ is
\begin{equation}
N_I(V_4)=\int_{V_4} d^4x \sqrt{g} D_I(x)~,
\label{eq-density}
\end{equation}
where $g=|\det g_{ab}|$.  (The special case of pointlike events can be recovered by writing $D_I$ as a sum of $\delta$-functions.) 

The infinitesimal number of events of type $I$ in the Short Fat Geodesic emitted from the point $x_0\in\Sigma_0$ is 
\begin{equation}
dN_I^{\rm FG}=dV \int_{\eta>0} d\tau  D_I(x(\tau))~,
\end{equation}
where $\tau$ is the proper time along the geodesic, and the integral is restricted to portions of the geodesic with positive scale factor parameter.  $dV$ is a fixed infinitesimal volume, which we may choose to define on $\Sigma_0$:
\begin{equation}
dV\equiv dV_0=d^3x_0 \sqrt{h_0}~.
\end{equation}
The total number of events in the ensemble of fat geodesics is obtained by integrating over all geodesics emanating from $\Sigma_0$:
\begin{equation}
N_I^{\rm FG}=\int_{\Sigma_0} d^3x_0 \sqrt{h_0} \int_{\eta>0} d\tau D_I(x(\tau))~.
\label{eq-nfg}
\end{equation}
where $h_0$ is the root of the determinant of the three-metric on $\Sigma_0$. We may take Eq.~(\ref{eq-nfg}) as the definition of the fat geodesic measure, with relative probabilities given by
\begin{equation}
\frac{P_I^{\rm FG}}{P_J^{\rm FG}}=\frac{N_I^{\rm FG}}{N_J^{\rm FG}}~.
\end{equation}

Note that Eq.~(\ref{eq-nfg}) is not a standard integral over a four volume; it is an integral over geodesics.  We may rewrite it as an integral over a four-volume because the definition of the Short Fat Geodesic ensures that the geodesics do not intersect.\footnote{Strictly, the definition only ensures that infinitesimally neighboring geodesics do not intersect.  We assume that $\Sigma_0$ is chosen so as to avoid nonlocal intersections (between geodesics with distinct starting points on $\Sigma_0$).  We expect that this is generic due to the inflationary expansion, and in particular that it is satisfied for $\Sigma_0$ in the attractor regime of the New Scale Factor Cutoff.  In structure forming regions, we expect that caustics occur before nonlocal intersections.  If not, then some regions may be multiply counted~\cite{Bou12b}; this would not affect the duality.} Then the coordinates $t, x_0$ define a coordinate system in the four-volume traced out by the congruence.  However, the four-volume element is {\em not\/} $d^3x_0 d\tau \sqrt{h_0} $.   Because the geodesics in the volume element $d^3x_0 \sqrt{h_0}$ expand along with the congruence, the correct four-volume element is 
\begin{equation}
d^4x \sqrt{g}=d^3x_0 d\tau \sqrt{h_0} e^{3\eta}~.
\end{equation}
This follows from the definition of expansion and scale factor parameter, Eqs.~(\ref{eq-sft}) and (\ref{eq-exp}).

\begin{figure}
\begin{center}
\includegraphics[scale = .8]{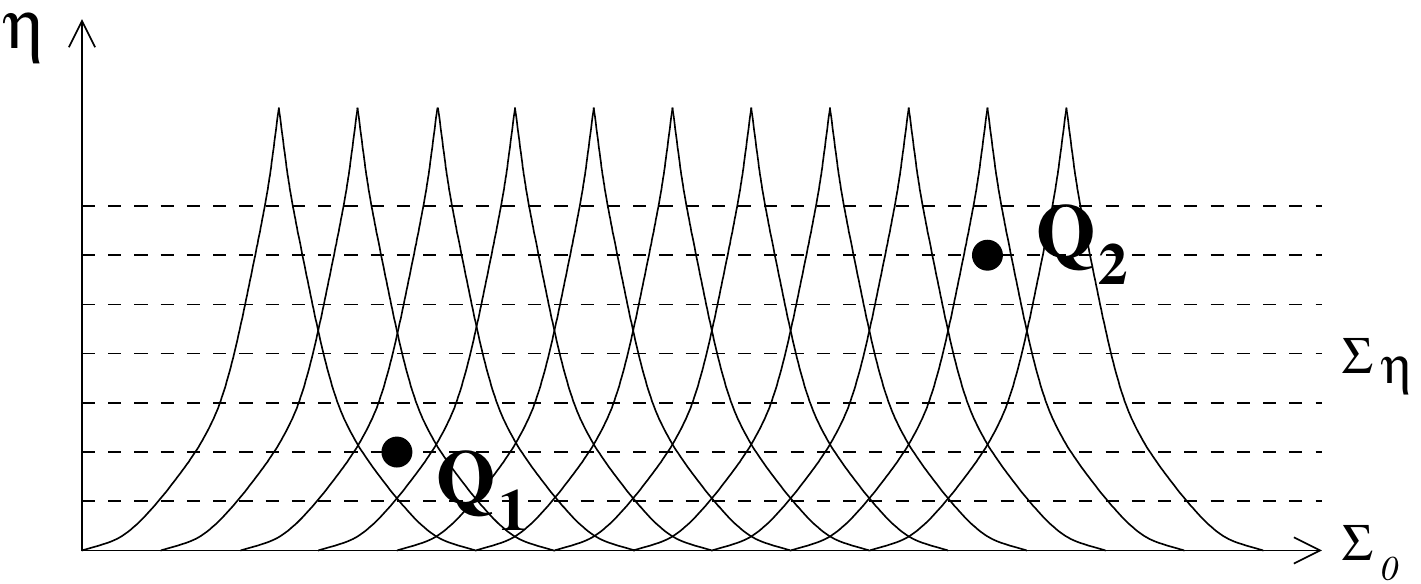}
\end{center}
\caption{Like the ensemble of causal patches in the previous section, the ensemble of fat geodesics probe the entire spacetime, but with a weighting (or overlap factor) that decreases exponentially with time.  In the discretized example shown, event $Q_1$ is double-counted whereas the later event $Q_2$ is counted only by one fat geodesic.  Because the weighting depends only on the scale factor time $\eta$, the fat geodesic cutoff is dual to the scale factor time cutoff if initial conditions for the former are chosen in the attractor regime of the latter.---Like the previous figure, this is a Penrose diagram.  The fat geodesics have fixed physical width but appear to be shrinking due to the conformal rescaling.}
\label{fig-fg}
\end{figure}

Returning to the event count, we can now write Eq.~(\ref{eq-nfg}) as an integral over the spacetime region $V_4(\eta>0)$ traced out by the congruence of Short Fat Geodesics:
\begin{equation}
N_I^{\rm FG}=\int_{V_4(\eta>0)} d^4y\sqrt{g(y)} e^{-3\eta} D_I(y)~.
\label{eq-fgd0}
\end{equation}
The weighting factor $e^{-3\eta}$ can be understood intuitively as the number of fat geodesics that overlap at each spacetime point; see Fig.~\ref{fig-fg}.

In particular, we may choose the scale factor parameter $\eta$ as a coordinate.  However, the coordinate is one-to-one only if we restrict to the expanding or the collapsing portion of each geodesic.  Thus, we may write
\begin{equation}
N_I^{\rm FG}=N_I^{{\rm FG},+}+N_I^{{\rm FG},-}~,
\label{eq-fgd}
\end{equation}
where
\begin{equation} 
N_I^{{\rm FG},\pm}=\int_0^\infty d\eta' e^{-3\eta'} \int_{\Sigma^\pm_{\eta'}} d^3x \sqrt{g(\eta',x)} D_I(\eta',x)~.
\label{eq-nifgpm}
\end{equation}
Here, $\Sigma^\pm_{\eta'}$ are hypersurfaces of constant scale factor parameter $\eta'$ in the expanding ($+$) or contracting ($-$) portion of the congruence.

\subsection{Proof of Equivalence to the New Scale Factor Cutoff}

We now turn to the global side of the duality.  Again, we consider the congruence of geodesics orthogonal to an initial surface $\Sigma_0$.  The New Scale Factor Cutoff measure is defined as
\begin{equation}
\frac{P_I^{\rm SF}}{P_J^{\rm SF}}=\lim_{\eta\to\infty} \frac{N_I(\eta)}{N_J(\eta)}~,
\label{eq-sfprob}
\end{equation}
where $N_I(\eta)$ is the number of events of type $I$ that have taken place in the spacetime regions with scale factor parameter less than $\eta$~\cite{Bou12b}.  Using Eq.~(\ref{eq-density}), we may write this as
\begin{equation}
N_I(\eta)=\int_{M(\eta)} d^4x \sqrt{g} D_I(x)~.
\label{eq-neta}
\end{equation}
The integral runs over the spacetime four-volume $M(\eta)$, defined as the set of points that lie in the future $\Sigma_0$ (on which we set $\eta=0$) and whose scale factor time, Eq.~(\ref{eq-sft}), is less than $\eta$.  In order to make the assignment of a scale factor parameter to every spacetime point unique, each geodesic is terminated immediately prior to caustic points, when neighboring geodesics intersect.

To compute the probabilities defined by the scale factor time cutoff, we note that the cosmological dynamics of eternal inflation leads to an attractor regime~\cite{Bou12b}:
\begin{equation}
N_I(\eta)=\bar N_I e^{\gamma \eta} + O(e^{\phi \eta})~,
\label{eq-sfattract}
\end{equation}
where $\phi<\gamma<3$.   This will be reviewed in the next subsection; for now, we will simply use this result.  With Eq.~(\ref{eq-sfprob}), it implies that the scale factor time probabilities are given by
\begin{equation}
\frac{P_I^{\rm SF}}{P_J^{\rm SF}}=\frac{\bar N_I}{\bar N_J}~.
\label{eq-sfprob2}
\end{equation}

The Short Fat Geodesic measure can also be evaluated using Eq.~(\ref{eq-sfattract}), if initial conditions on $\Sigma_0$ are chosen to lie in the attractor regime.  A suitable $\Sigma_0$ can be constructed as as a late-time hypersurface orthogonal to the congruence constructed from a much earlier, arbitrary initial hypersurface, and resetting $\eta\to 0$ there. The proof will exploit the fact that the Short Fat Geodesic probabilities, Eqs.~(\ref{eq-fgd}) and (\ref{eq-nifgpm}), involve an integral over a spacetime set closely related to $M(\eta)$, reweighted relative to Eq.~(\ref{eq-neta}) by a factor that depends only on $\eta$ and thus does not change relative probabilities in the attractor regime.

The set $M(\eta)$ will contain one connected expanding region, $M^+(\eta)$, bounded from below by $\Sigma_0$, in which scale factor time is growing towards the future.  In any model with collapsing regions (structure forming regions or crunches), $M(\eta)$ will also contain infinitely many mutually disconnected collapsing regions inside bubbles near the future conformal boundary of the spacetime.  (The total contribution to the measure from such regions to the New Scale Factor Cutoff measure is finite at any finite value of $\eta$~\cite{Bou12b}.)  We denote the union of all collapsing regions by $M^-(\eta)$.  

Let us split the integral in Eq.~(\ref{eq-neta}) into expanding and contracting portions:
\begin{equation}
N_I(\eta)=N_I^+(\eta)+N_I^-(\eta)~,
\label{eq-netapm}
\end{equation}
where 
\begin{equation}
N^{\pm}_I(\eta)\equiv \int_{M^\pm(\eta)} d^4x \sqrt{g} D_I(x) ~.
\label{eq-netapm2}
\end{equation}
Since this division depends only on local properties, each portion has its own attractor solution:
\begin{equation}
N^\pm_I(\eta)=\bar{N}^\pm_I e^{\gamma \eta} + O(e^{\phi \eta})~,
\label{eq-sfattract-pm-dual}
\end{equation}
with $\bar{N}^{+}_I+\bar{N}^{-}_I=\bar{N}_I$. This will be shown explicitly in the following subsection.
In each portion, the scale factor parameter is monotonic along the geodesics in the congruence, and we may use it as an integration variable:
\begin{eqnarray} 
N^{+}_I(\eta) & = & 
\int_0^\eta d\eta' \int_{\Sigma^{+}_{\eta'}} d^3x \sqrt{g(\eta',x)} D_I(\eta',x)~;\\
N^{-}_I(\eta) & = & 
\int_{-\infty}^\eta d\eta' \int_{\Sigma^{-}_{\eta'}} d^3x \sqrt{g(\eta',x)} D_I(\eta',x)~,
\end{eqnarray} 
where the hypersurface $\Sigma^\pm_{\eta'}$ consists of the points with fixed $\eta'$ in the expanding ($+$) or contracting ($-$) region.  Therefore
\begin{equation}
\frac{dN^\pm_I}{d\eta}=  \int_{\Sigma^\pm_\eta} d^3x \sqrt{g(\eta,x)} D_I(\eta,x)~.
\label{eq-dndeta}
\end{equation}
so we may rewrite Eq.~(\ref{eq-nifgpm}) as
\begin{equation}
N_I^{{\rm FG},\pm}=\int_0^\infty d\eta' e^{-3\eta'} \frac{dN^\pm_I}{d\eta'}~.
\label{eq-nifgpm2}
\end{equation}
We can now use the attractor solutions for the New Scale Factor Cutoff, Eq.~(\ref{eq-sfattract-pm-dual}), to evaluate the Short Fat Geodesic cutoff:
\begin{eqnarray} 
N_I^{\rm FG} & = & \int_0^\infty d\eta' e^{-3\eta'} \left(\frac{dN^{+}_I}{d\eta'}+\frac{dN^{-}_I}{d\eta'}\right)\\
 & = & \int_0^\infty d\eta' e^{(\gamma-3)\eta'}\gamma \bar N_I \\
 & = & \frac{\gamma}{3-\gamma} \bar N_I~.
\label{eq-fgd2}
\end{eqnarray} 
The prefactor is $I$-independent, so relative probabilities in the Short Fat Geodesic measure are given by
\begin{equation}
\frac{P_I^{\rm FG}}{P_J^{\rm FG}}=\frac{\bar N_I}{\bar N_J}~.
\end{equation}
This agrees with the New Scale Factor Cutoff probabilities, Eq.~(\ref{eq-sfprob2}).  Therefore, the two measures are equivalent.

This result is somewhat counterintuitive. If the Short Fat Geodesic is defined for regions where $\eta>0$, its global dual should be the New Scale Factor Cutoff not as defined above, but restricted to $\eta>0$ both in the expanding and collapsing regions.  In fact, it is.  Both versions of the New Scale Factor Cutoff, with and without this additional restriction in the collapsing regions, are dual to the Short Fat Geodesic, because both have the same attractor regime. It is important to distintuish between $\eta<0$ regions and decreasing-$\eta$ regions in the global cutoff. The former quickly become unimportant in the attractor regime; the latter are always important. For the cumulative quantity $N_I(\eta)$ to be exactly in the attractor regime, it would be necessary to restrict to $\eta>0$ on both the expanding and collapsing side, and thus to exclude a few collapsing regions; otherwise, there may be a small transient of order $e^{\phi\eta}$ from those initial collapsing regions that have $\eta<0$.  However, the duality relies on evaluating the local measure by integrating up the global ``derivative" $dN_I/d\eta$ with weighting $e^{-3\eta}$.  Since the calculation makes reference only to the derivative in the region $\eta>0$, the regions with $\eta<0$ do not enter into the duality.

\subsection{New Scale Factor Cutoff Rate Equation and Attractor Solution}
\label{sec-sfattract}

In this subsection we derive the attractor solution,  Eq.~(\ref{eq-sfattract}),  starting from the rate equation for the New Scale Factor Cutoff.  We will follow~\cite{Bou12b} and use the seminal results of~\cite{GarSch05}.  As for the case of light-cone time, we will proceed in two steps.  We first consider the rate equation for de~Sitter vacua; then we include the detailed consequences of decays within this distribution, and explain how to treat collapsing regions.  

Naively, the rate equation should follow from the result for light-cone time, Eq.~(\ref{eq-lcattract}), by an appropriate substitution.   In empty de~Sitter space, $\theta/3=\tau_{\Lambda,\alpha}^{-1}$, so Eqs.~(\ref{eq-sft}) and (\ref{eq-dtlcdt}) imply that $d t_{\rm LC} =d\eta$ in this regime.  Setting $d t_{\rm c} \to d\eta$ in Eq.~(\ref{eq-lcattract}), however, yields an incorrect equation:
\begin{equation}
\frac{dn_\alpha}{d\eta} = (3-\kappa_\alpha) n_\alpha + \sum_\beta \kappa_{\alpha \beta} n_\beta~~~?
\label{eq-nratewrong}
\end{equation}
The last term on the right hand side is incorrect.  In the light-cone time rate equation, this term arose from the square bubble approximation.  It is an effective term that anticipates the asymptotic size of bubbles of new vacua instead of treating their growth in detail.  It subsumes, in particular, the cumulative effects of the early era within a new bubble (less than $\tau_{\Lambda,\alpha}$ after nucleation). During this era the relation $d t_{\rm LC} = d\eta$ does {\em not\/} hold, so the substitution that led to Eq.~(\ref{eq-nratewrong}) is unjustified.  Another way of saying this is that the square bubble approximation is a different procedure for different time variables.

The correct rate equation for the New Scale Factor Cutoff contains an extra factor of $v_\alpha/v_\beta$ in the final sum, where $v_\alpha$ is the proper volume of a horizon patch of type $\alpha$.  It thus takes a particularly simple form, 
\begin{equation}
\frac{dV_\alpha}{d\eta} = (3-\kappa_\alpha) V_\alpha + \sum_\beta \kappa_{\alpha \beta} V_\beta~,
\label{eq-vrate}
\end{equation}
when expressed in terms of the proper volumes $V_\alpha$ occupied by metastable de~Sitter vacua $\alpha$ at scale factor time $\eta$, instead of the number of horizon patches $n_\alpha=V_\alpha/v_\alpha$.  More generally, one finds that the rate equation takes the above form, with $V\to X$, $\eta\to T$, if $T$ measures the growth of the overall volume of space in units of $X$.  For example, scale factor parameter measures the growth of proper volume ($T=\eta$, $X=1$) and light-cone time measures the growth of volume in units of horizon volume ($T= t_{\rm LC}$, $X=v_\alpha$).

To derive Eq.~(\ref{eq-vrate}), we note that the first two terms on the right hand side follow from the arguments given for the analogous terms in Sec.~\ref{sec-sfattract}.  They would also follow from Eq.~(\ref{eq-nrate}) by substituting $d t_{\rm LC} \to d\eta$ and using $V_\alpha=n_\alpha v_\alpha$; but the third term, as explained above, cannot be so obtained.  It must be derived from a first principle argument identical to that given in Sec.~\ref{sec-sfattract}; except that it is now the continuity of {\em New Scale Factor parameter}, not light-cone time, that must be ensured when a new bubble is formed.  This means that instead of requiring that the number of horizon patches of $\beta$-vacuum lost must equal the number of horizon patches of $\alpha$-vacuum gained in $\beta\to \alpha$ transitions, we now require that the proper volume of $\beta$-vacuum lost must equal the proper volume of $\beta$-vacuum gained.  The amount lost in $\beta$, per nucleation of $\alpha$, is always one horizon volume of $\beta$; this follows from causality.  In the rate equation in terms of scale factor parameter, this must be converted into proper volume, and the same proper volume must be assigned to the new vacuum, $\alpha$.   This leads to the final term in Eq.~(\ref{eq-vrate}).  It also explains why the New Scale Factor rate equation looks simplest in terms of proper volume.

The solution of the rate equation for New Scale Factor time can be obtained from Eq.~(\ref{eq-nsolution}) by substituting  $t_{\rm LC} \to \eta$ and $n\to V$:
\begin{equation}
V_\alpha(\eta)=\bar V_{\alpha} e^{\gamma\eta}+O(e^{\varphi\eta})~.
\end{equation}
That is, we must set $\bar V_\alpha$ (and not, as for light-cone time, $\check n_\alpha$), equal to the dominant eigenvector of the transition matrix $M_{\alpha \beta}$.   As before $\gamma=3-q$ is the largest eigenvalue. Note that this eigenvector and the dominant vacuum $*$ are exactly the same as in the case of light-cone time.  On a slice of constant light-cone time, the $*$ vacuum dominates the number of horizon patches; on a slice of constant New Scale Factor Cutoff, it dominates the volume.

It is convenient to define 
\begin{equation}
\bar n_\alpha\equiv \frac{\bar V_\alpha}{v_\alpha}~.
\label{eq-barni1}
\end{equation}
The number of horizon patches of type $\alpha$ at New Scale Factor parameter $\eta$ obeys
\begin{equation}
n_\alpha(\eta)=\bar n_\alpha e^{\gamma\eta}+O(e^{\varphi\eta})~.
\label{eq-nibarni}
\end{equation}

We now derive $N_I(\eta)$. The procedure will be slightly different from the one in Ref.~\cite{Bou12b} in that we will keep expanding and collapsing regions explicitly separated in all expressions.  $N_I(\eta)$ receives a contribution from the expanding (+) regions, and one from the contracting (-) regions, $N_I(\eta)=N^{+}_I(\eta)+N^{-}_I(\eta)$. Analogous to Eq.~(\ref{eq-nilct1}) for the Lightcone Time Cutoff, one finds for the New Scale Factor Cutoff:
\begin{equation}
N^{+}_I(\eta)=\kappa_{I*} n_*(\eta)+\sum_{i\neq *}\int_0^{\eta}
\left(\frac{dN^{+}_I}{dN_i}\right)_{\eta-\eta_{\rm nuc}} 
\left(\frac{dN_i}{d \eta }\right)_{\eta_{\rm nuc}} d \eta_{\rm nuc} ~,
\label{eq-nisft1-+new}
\end{equation}
and
\begin{equation}
N^{-}_I(\eta)=\sum_{i\neq *}\int_{-\infty}^{\eta}
\left(\frac{dN^{-}_I}{dN_i}\right)_{\eta-\eta_{\rm nuc}} 
\left(\frac{dN_i}{d \eta }\right)_{\eta_{\rm nuc}} d \eta_{\rm nuc} ~.
\label{eq-nisft1--new}
\end{equation}
As in the previous section, we can now change the integration variable to $\zeta=\eta-\eta_{\rm nuc}$ in Eqs. (\ref{eq-nisft1-+new}) and (\ref{eq-nisft1--new}), and use (\ref{eq-nibarni}) to get
\begin{equation}
N^\pm_I(\eta)=\bar{N}^\pm_I e^{\gamma \eta} + O(e^{\phi \eta})~,
\label{eq-sfattract-pm-dual-bis}
\end{equation}
where
\begin{equation}
\bar N^+_I\equiv \kappa_{I*} \bar n_*+\sum_{i\neq *}\sum_\beta N^+_{Ii} \kappa_{i\beta} \bar n_\beta~,
\label{eq-barni-new+}
\end{equation}
and
\begin{equation}
\bar N^-_I\equiv \sum_{i\neq *}\sum_\beta N^-_{Ii} \kappa_{i\beta} \bar n_\beta~,
\label{eq-barni-new-}
\end{equation}
with
\begin{equation}
N^{+}_{Ii}\equiv \int_{0}^\infty d\zeta e^{-\gamma \zeta} \left( \frac{dN^{+}_I}{dN_i}\right)_{\zeta}~,
\label{eq-nii2+new}
\end{equation}
and
\begin{equation}
N^{-}_{Ii}\equiv \int_{-\infty}^\infty d\zeta e^{-\gamma \zeta} \left( \frac{dN^{-}_I}{dN_i}\right)_{\zeta}~.
\label{eq-nii2-new}
\end{equation}
Again, (\ref{eq-nii2+new}) and (\ref{eq-nii2-new}) converge because all vacua decay faster than the dominant vacuum. This yields Eq. (\ref{eq-sfattract}), with $\bar{N}_I=\bar{N}^+_I+\bar{N}^-_I$.

\section{General Global-Local Dualities}
\label{sec-x}

It is easy to generalize the duality studied in the previous section, between the Short Fat Geodesic cutoff and the New Scale Factor Cutoff.  On the local side, the fatness of the geodesic can be allowed to vary along the geodesic.   On the global side, this corresponds to a different choice of time variable.   In this section, we mostly consider a generalization that preserves the key feature that the fatness of the geodesic does not depend explicitly on the time along the geodesic, but only on local features.   This restriction defines a family of measures that include the scale factor cutoff, the light-cone time cutoff, and the CAH cutoff as special cases.  In the final subsection, we consider a further generalization that we exemplify by deriving a local dual to the proper time cutoff.

\subsection{$X$-fat Geodesic Measure}

Consider a family of geodesics orthogonal to an initial hypersurface $\Sigma_0$. We assign each of these geodesics a cross sectional volume, $X$. Intuitively, we may picture $X$ as modulating the infinitesimal fatness of the geodesic.  Equivalently, $X$ can be thought of as a weighting factor that allows events of type $I$ to contribute differently to the probability for $I$, depending on where they are encountered. We require that $X$ be everywhere nonnegative to ensure that probabilities are nonnegative. We will assume, for now, that $X$ depends only on local properties of the congruence, such as the expansion, the shear,  and their derivatives:
\begin{equation}
X=X(\theta, \sigma^{ab}\sigma_{ab}, \frac{d\theta}{d\tau}, \ldots)~.
\end{equation}
A simple example, which we will consider explicitly in Sec.~\ref{sec-hubbletube}, is the {\em Hubbletube}.  It is obtained by setting $X$ to the local Hubble volume.  With $X\equiv 1$, the $X$-fat geodesic reduces to the ordinary fat geodesic.  In Sec.~\ref{sec-propertime}, we will consider further generalizations, in which $X$ is not restricted to a local function of congruence parameters. 

Along each geodesic, we may define the {\em T parameter}: 
\begin{equation}
T\equiv \eta-\frac{1}{3}\log X
\label{eq-tx}
\end{equation}
Geometrically, $e^{3T}$ is the factor by which a volume element has expanded along the congruence, in units of the volume $X$. Every {\em$X$-Fat Geodesic} will be restricted to values of the $T$ parameter larger than that at $\Sigma_0$, which we may choose to be zero.\footnote{Strictly, this should be called the {\em short} $X$-fat geodesic: as in Sec.~\ref{sec-fgdef}, we will be restricting the congruence to regions where its density (in units of the local fatness, $X$) is below its initial value. This ensures the broadest possible applicability of the duality we derive.  By including the entire future-directed geodesic irrespective of this conditions, one could consider a ``long'' $X$-fat geodesic.  This local measure would not generally have a natural global dual.} 

We would like to compute probabilities using our new local cut-off, the $X$-fat geodesic, by modifying the definition of the fat geodesic, Eq.~(\ref{eq-nfg}):
\begin{equation}
N_I^{\rm XG}=\int_{\Sigma_0} d^3x_0 \sqrt{h_0} \int_{T>0} d\tau X(x(\tau)) D_I(x(\tau))~.
\label{eq-nxg}
\end{equation}
Relative probabilities given by
\begin{equation}
\frac{P_I^{\rm XG}}{P_J^{\rm XG}}=\frac{N_I^{\rm XG}}{N_J^{\rm XG}}~.
\label{rel-prop-beg}
\end{equation}
We now follow the steps leading to Eq.~(\ref{eq-fgd0}) in Sec.~\ref{sec-fgdef}.  Assuming that the geodesics do not intersect, Eq.~(\ref{eq-nxg}) can be rewritten as an standard integral over the four-volume encountered by the congruence:
\begin{equation}
N_I^{\rm XG}=\int_{V_4(T>0)} d^4y\sqrt{g(y)} e^{-3\eta} X(y) D_I(y)~.
\label{eq-xgd0}
\end{equation}
In particular, we can pick $T$ as a coordinate. Like the scale factor time, $T$ in Eq.~(\ref{eq-tx}) is defined for every point on the nonintersecting congruence. Multiple points along the same geodesic may have the same $T$; this will not be a problem. However, the coordinate is one-to-one only if we restrict to the ``expanding" or the ``contracting" portion of each geodesic.\footnote{In this section, ``expanding" and ``contracting" regions are defined with respect to the $X$ volume. An expanding/contracting region will be one where $T$ increases/decreases.} Thus, in terms of $T$, Eq.~(\ref{eq-xgd0}) becomes 
\begin{equation}
N_I^{\rm XG}=N_I^{{\rm XG},+}+N_I^{{\rm XG},-}~,
\label{eq-xgd-sum}
\end{equation}
where
\begin{equation}
N_I^{\rm XG, \pm}=\int_0^\infty dT' e^{-3T'} \int_{\Sigma^\pm_{T'}} d^3x \sqrt{g(T',x)} D_I(T',x)~,
\label{eq-xgd}
\end{equation}

Here, $\Sigma^\pm_{T'}$ are hypersurfaces of constant $T$ parameter $T'$ in the expanding ($+$) or contracting ($-$) portion of the congruence.

\subsection{Proof of Equivalence to the $T$-cutoff Measure}

Let us consider the time variable $T$ defined in Eq.~(\ref{eq-tx}) as a global cutoff.  Probabilities are defined by
\begin{equation}
\frac{P_I^T}{P_J^T}=\lim_{T\to\infty} \frac{N_I(T)}{N_J(T)}~,
\label{eq-Tprob}
\end{equation}
where $N_I(T)$ is the number of events of type $I$ that take place in spacetime regions with time less than $T$.  Because $T$ need not be monotonic along every geodesic, such regions may not be connected. As shown in Sec. \ref{sec-sfattract}, this does not affect the proof of equivalence, which proceeds as in Sec.~\ref{sec-fg}. Again, $N_I(T)$ receives a contribution from expanding (+) and contracting (-) regions, $N_I(T)=N_I^{+}(T)+N_I^{-}(T)$. 

In terms of the distribution $D$, 
\begin{eqnarray} 
N^{+}_I(T) & = & 
\int_0^T dT' \int_{\Sigma^{+}_{T'}} d^3x \sqrt{g(T',x)} D_I(T',x)~;\\
N^{-}_I(\eta) & = & 
\int_{-\infty}^T dT' \int_{\Sigma^{-}_{T'}} d^3x \sqrt{g(T',x)} D_I(T',x)~,
\end{eqnarray} 
and therefore
\begin{equation}
\frac{dN^\pm_{I}}{dT}=  \int_{\Sigma^\pm_{T}} d^3x \sqrt{g(T,x)} D_I(T,x)~.
\label{eq-dndT}
\end{equation}
We make use of the attractor solution 
\begin{equation}
N_I(T)=\tilde N_I e^{\gamma T} + O(e^{\phi T})~,
\label{eq-Tattract}
\end{equation}
where $\phi<\gamma<3$ (see the following subsection).  With Eq.~(\ref{eq-Tprob}), it implies that the $T$-cutoff probabilities are given by
\begin{equation}
\frac{P_I^T}{P_J^T}=\frac{\tilde N_I}{\tilde N_J}~.
\label{eq-Tprob2}
\end{equation}

The $X$-fat geodesic probabilities are also determined by Eq.~(\ref{eq-Tattract}), {\em if\/} initial conditions on $\Sigma_0$ are chosen to lie in the attractor regime. Plugging  Eq. ~(\ref{eq-dndT}) into Eq. ~(\ref{eq-xgd}), and then using Eq.~(\ref{eq-xgd-sum}), we get
\begin{equation}
N_I^{\rm XG}=\int_0^\infty dT' e^{-3T'} \frac{dN_I}{dT'}~,
\label{eq-xgd2}
\end{equation}
In the attractor regime, by Eq.~(\ref{eq-Tattract}), one has
\begin{equation}
\frac{dN_I}{dT'}=\gamma \tilde N_I e^{\gamma T'}~.
\end{equation}
Substituting into Eq.~(\ref{eq-xgd2}) and using $\gamma<3$, the integral converges to an $I$-independent constant.  Thus, relative probabilities in the $X$-fat geodesic measure are given by
\begin{equation}
\frac{P_I^{\rm XG}}{P_J^{\rm XG}}=\frac{\tilde N_I}{\tilde N_J}~.
\end{equation}
This agrees with the $T$-cutoff probabilities, Eq.~(\ref{eq-Tprob2}).  Therefore, the two measures are equivalent.
 
\subsection{$T$-cutoff Rate Equation and Attractor Solution}

In this subsection, we derive the rate equation for the number of horizon patches of de~Sitter vacua $\alpha$ as a function of $T$, and the attractor solution, Eq.~(\ref{eq-Tattract}).  For the rate equation, we treat all de~Sitter vacua as empty at all times.  We use the square bubble approximation which treats each bubble as comoving in the congruence at its asymptotic size.  

Let $x_\alpha$ be the asymptotic value of $X$ in the vacuum $\alpha$.  $X$ will converge rapidly to $x_\alpha$ in empty de~Sitter regions because, by assumption, $X$ depends only on local properties of the congruence.  By Eq.~(\ref{eq-tx}) this implies that $dT=d\eta$ in such regions.  Thus, the rate equation is
\begin{equation}
\frac{dn_\alpha}{dT} =(3-\kappa_\alpha) n_\alpha + \sum_\beta \kappa_{\alpha \beta} n_\beta  \frac{v_\beta x_\alpha}{v_\alpha x_\beta}~.
\end{equation}
The term $3n_\alpha$ captures the exponential growth of the number of horizon patches, which goes as $e^{3\eta}$.  The term $-\kappa_\alpha n_\alpha$ captures the decay of vacuum $\alpha$, per unit horizon patch and unit scale factor time in empty de~Sitter space.

The last term captures the creation of new regions of vacuum $\alpha$ by the decay of other vacua.  In the square bubble approximation, one horizon patch of $\beta$ is lost when an $\alpha$-bubble forms in $\beta$ (see Sec.~\ref{sec-lcattract}).  Thus, $v_\beta/x_\beta$ $X$-patches of $\beta$-vacuum are lost, where $v_\beta$ is the volume of one horizon patch of $\beta$.  Continuity of the time variable $T$ requires that the number of patches of size $X$ be continuous, so $v_\beta/x_\beta$ $X$-patches of $\alpha$ vacuum must be added.  One $X$-patch of $\alpha$ vacuum equals $x_\alpha/v_\alpha$ horizon patches of $\alpha$ vacuum.  Thus, the total number of horizon patches of $\alpha$-vacuum that are created per $\beta$-decay in the square bubble approximation is $\frac{v_\beta x_\alpha}{v_\alpha x_\beta}$.
The number of such decays in the time interval $dT$ is $\kappa_{\alpha \beta} n_\beta dT$.  This completes our derivation of the last term.

When expressed in terms of the number of $X$-patches, 
\begin{equation}
n^X_\alpha=\frac{n_\alpha v_\alpha}{x_\alpha} = \frac{V_\alpha}{x_\alpha}~,
\label{eq-xpatch}
\end{equation}
the rate equation takes a very simple form:
\begin{equation}
\frac{dn^X_\alpha}{dT} =(3-\kappa_\alpha) n^X_\alpha + \sum_\beta \kappa_{\alpha \beta} n^X_\beta ~.
\label{eq-xrate}
\end{equation}
This form is identical to that of Eqs.~(\ref{eq-nrate}) and (\ref{eq-vrate}), and the general results of Ref.~\cite{GarSch05} apply.   The late-time solution is again determined by the dominant eigenvalue, $\gamma=3-q$, of the transition matrix $M_{\alpha \beta}$, and by the associated eigenvector, which we now label $\tilde n^{X}_\alpha$:
\begin{equation}
n^X_\alpha(T)=\tilde n^X_\alpha e^{\gamma T} + O(e^{\varphi T})~.
\label{eq-nsolution-T}
\end{equation}

Next, we compute number of events of type $I$ prior to the time $T$.  With
\begin{equation}
\tilde n_\alpha\equiv \frac{\tilde n^X_\alpha x_\alpha}{v_\alpha}~,
\label{eq-tildeni2}
\end{equation}
the number of horizon patches of type $\alpha$ at time $T$ obeys
\begin{equation}
n_\alpha(T)=\tilde n_\alpha e^{\gamma T}+O(e^{\varphi T})~.
\label{eq-nibarniT}
\end{equation}
The remainder of the analysis is completely analogous to Sec.~\ref{sec-sfattract}. When we include collapsing (i.e. decreasing $T$) regions at the future of $\Sigma_0$, we still obtain an attractor regime.  Like in the New Scale Factor case \cite{Bou12b}, the corresponding  $\Delta T_-=T_{\rm{max}}-T_{\dag}$\footnote{Here, $T_{\dag}$ corresponds to the $T$ value at the regulated endpoint of the geodesic. Geodesics are terminated at some cutoff, for example one Planck time before they reach a point where $T\to -\infty$. The choice of cutoff depends on the definition of $T$; see the discussion in the next subsection. On the other hand, $T_{\rm{max}}$ is the maximum $T$-value reached by the geodesic.} and $\Delta T_+=T_{\rm{max}}-T_{\rm{nuc}}$ will only depend on local physics in each bubble universe but not on $T_{\rm{nuc}}$. This holds because we are assuming that $X$ only depends on local properties of the congruence. Therefore, $\Delta T_-$ and $\Delta T_+$ will increase by the same finite amounts during expansion and collapse phases in a particular pocket universe, no matter when the bubble universe is nucleated.  There will be infinitely many collapsing regions at the future of $\Sigma_0$, but a finite number of bubbles contribute, namely the ones that formed before the time $T+\Delta T_{\rm{sup}}$, where $\Delta T_{\rm sup}\equiv\min\{0,\sup_{x_0} (\Delta T_--\Delta T_+)\}$

\begin{equation}
N_I(T)=\kappa_{I*} n_*+\sum_{i\neq *}\int_0^{T+\Delta T_{\rm sup}}
\left(\frac{dN_I}{dN_i}\right)_{T-T_{\rm nuc}} 
\left(\frac{dN_i}{d T }\right)_{T_{\rm nuc}} d T_{\rm nuc} ~.
\label{eq-nisft1T}
\end{equation}
Again, we conclude
\begin{equation}
N_I(T)=\bar N_I e^{\gamma T} + O(e^{\phi T})~,
\label{eq-Tattract1-end}
\end{equation}
where
\begin{equation}
\bar N_I\equiv \kappa_{I*} \bar n_*+\sum_{i\neq *}\sum_\beta N_{Ii} \kappa_{i\beta} \bar n_\beta~,
\label{eq-barni-T}
\end{equation}
and
\begin{equation}
N_{Ii}\equiv \int_{-\Delta T_{\rm sup}}^\infty d\zeta e^{-\gamma \zeta} \left( \frac{dN_I}{dN_i}\right)_{\zeta}
\label{eq-nii2-T}
\end{equation}
As in \cite{Bou12b}, this integral converges because all vacua decay faster than the dominant vacuum, and one obtains the same attractor behavior.

\subsection{The Hubbletube and the CAH measure}
\label{sec-hubbletube}

An example of particular interest is the {\em Hubbletube}: the $X$-fat geodesic whose fatness is proportional to the local Hubble volume $v_H$, as measured by the expansion of the congruence:
\begin{equation}
X\propto v_H=\frac{4\pi}{3} (\frac{3}{\theta})^3~.
\end{equation}
Since constant numerical factors drop out of all relative probabilities, we simply set
\begin{equation}
X\equiv \theta^{-3}~.
\end{equation}
This measure is dual to a global cutoff at constant $T$, where
\begin{equation}
T\equiv \eta+\log\theta~.
\end{equation}
Equivalently, the global cutoff surfaces can be specified in terms of any monotonic function of $T$, e.g. $\exp(T)$.  Note that 
\begin{equation}
e^T=\theta a~=\frac{da}{d\tau},
\end{equation}
where $a\equiv e^{\eta}$ is the scale factor and $\tau$ is proper time along the congruence.  We thus recognize the global dual of the Hubbletube as Vilenkin's CAH-cutoff~\cite{Vil11}.

Naively, the CAH-cutoff is well-defined only in regions with accelerating expansion: $\ddot a>0$, where the time variable $T$ increases monotonically along the geodesics.  In this regime, the duality with the Hubbletube is obvious.  But this regime is also extremely restrictive: it excludes not only gravitationally bound regions such as our galaxy, but also all regions in which the expansion is locally decelerating, including the homogeneous radiation and matter-dominated eras after the end of inflation in our vacuum.  

However, if geodesics are terminated before caustics, the CAH cutoff can instead be defined as a restriction to a set of spacetime points with $T$ less than the cutoff value. This is similar to the transition from the old to the new scale factor measure: in the spirit of Ref.~\cite{Bou12b}, one abandons the notion of $T$ as a time variable.  In the case of the CAH parameter $T$, an infinite number of decelerating regions will be included under the cutoff for any finite $T$.

This possibility of increasing the regime of applicability of the CAH cutoff is particularly obvious from the local viewpoint.   The local measure requires only $\theta>0$ for positive fatness; this is strictly weaker than $\ddot a>0$.  It still excludes collapsing regions, but not regions undergoing decelerating expansion.    

On either side of the duality, geodesics must be terminated at some arbitrarily small but finite proper time before they reach turnaround ($\theta=0$), where $T\to-\infty$.  Otherwise, events at the turnaround time receive infinite weight. This is needed only for finiteness; it eliminates an arbitrarily small region near the turnaround from consideration but does not affect other relative probabilities. However, this marks an important difference to the Short Fat Geodesic and the New Scale Factor measure, where no additional cutoff near $\eta\to-\infty$ was needed. 

In any case, the restriction to regions with $\theta>0$ is necessary to make the Hubbletube well-defined.  Unfortunately, this restriction is too strong to yield a useful measure since it excludes gravitationally bound regions like our own.  Unlike in the case of the New Scale Factor Cutoff or the Causal Patch, there are thus large classes of regions to which the CAH cutoff cannot be applied.  Additional rules must be specified, such as the CAH+ measure of Ref.~\cite{Vil11}.

\subsection{The Incredible Shrinking Geodesic and the Proper Time Cutoff}
\label{sec-propertime}

The global proper time cutoff is defined as a set of points that lie on a geodesic from $\Sigma_0$ with proper length (time duration) less than $\tau$ along the geodesic~\cite{Lin86a,LinLin94,GarLin94,GarLin94a,GarLin95}.  (To make this well-defined, we terminate geodesics at the first caustic as usual, so that every point lies on only one geodesic.)  Relative probabilities are then defined as usual, in the limit as the cutoff is taken to infinity.  

The rate equation for the number of de~Sitter horizon patches, in terms of proper time, is
\begin{equation}
\frac{dn_\alpha}{d\tau} = (3-\kappa_\alpha) H_\alpha n_\alpha + \sum_\beta \kappa_{\alpha \beta} H_\beta n_\beta = M_{\alpha\beta} n_\beta~,
\label{eq-nrateproper}
\end{equation}
where the transition matrix is given by
\begin{equation}
M_{\alpha\beta}=(3-\kappa_\alpha) H_\alpha \delta_{\alpha\beta} + \kappa_{\alpha \beta} H_\beta~.
\end{equation}
This differs from the transition matrix in all previous examples by the appearence of the Hubble constants of the de~Sitter vacua, and so it will not have the same eigenvector and eigenvalues; it will have a completely different attractor regime.  Instead of Planck units, it will be convenient to work in units of the largest Hubble constant in the landscape, $H_\alpha\to H_\alpha/H_{\rm max}$ and $\tau\to H_{\rm max}\tau$.  We note that $H_{\rm max}^{-1}$ is necessarily a microscopic timescale in any model where our vacuum contains a parent vacuum whose decay is sufficient for a reheat temperature consistent with nucleosynthesis.  In the string landscape, one expects $H_{\rm max}$ to be of order the Planck scale.  

Due to the smallness of decay rates and the large differences in the Hubble rate between Planck-scale vacua ($H\sim 1$) and anthropic vacua ($H\ll 1$), we expect that again the largest eigenvalue is very close to the largest diagonal entry in the transition matrix, and that the associated eigenvector is dominated by the corresponding vacuum.  In all previous examples, the dominant vacuum, $*$, was the longest-lived de~Sitter vacuum.  The associated eigenvalue was $\gamma\equiv 3-\kappa_*$, where $\kappa_*$ is the decay rate of the $*$ vacuum.  Now, however, the Hubble constant of each de~Sitter vacuum, $H_\alpha$, is the more important factor.  The dominant vacuum, $*$, will be the fastest-expanding vacuum, i.e., the vacuum with the largest Hubble constant, which in our unit conventions is $H_{\rm max}=1$.\footnote{More precisely, the dominant vacuum will be the vacuum with largest $(3-\kappa_\alpha)H_\alpha$.} In the same units, the associated eigenvalue is again $\gamma=3-\kappa_*$.  

By decay chains, the dominant expansion rate $\gamma$ drives both the growth rate of all other vacuum bubbles and all types of events, $I$, at late times:
\begin{equation}
N_I=\bar N_I e^{\gamma \tau} +O(e^{\varphi \tau})
\end{equation}
where $\varphi<\gamma<3$.  Relative probabilities are given as usual by
\begin{equation}
\frac{P_I}{P_J}=\frac{\bar N_I}{\bar N_J}~.
\end{equation}

\begin{figure}
\begin{center}
\includegraphics[scale = .5]{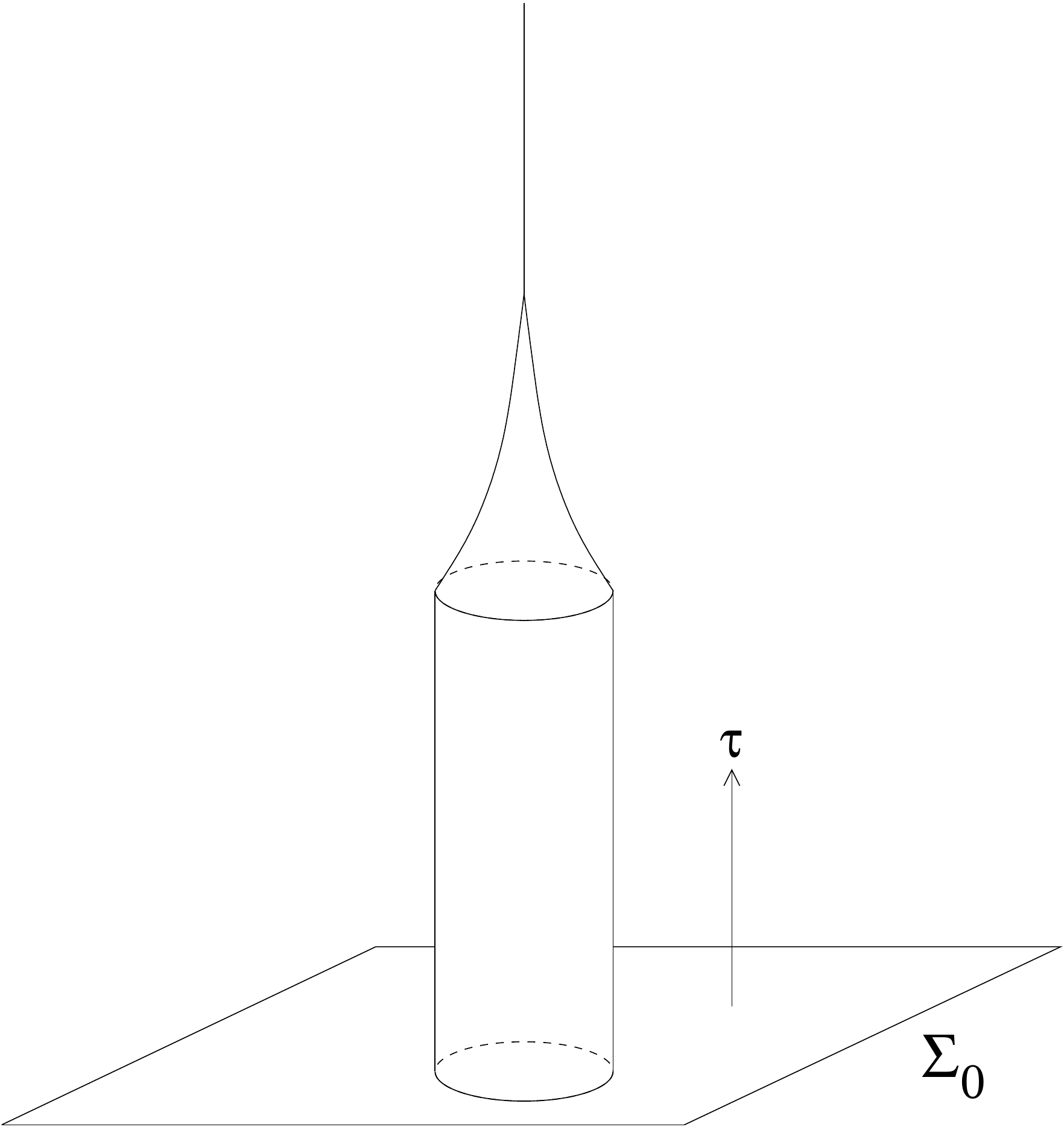}
\end{center}
\caption{The Incredible Shrinking Geodesic.  This is {\em not} a conformal diagram; the true proper fatness of the geodesic is shown as a function of proper time, $\tau$.  As long as the geodesic remains in the dominant vacuum, its fatness is constant, i.e., it assigns the same weight to all events it encounters.  In any other region, its fatness decreases exponentially with microscopic characteristic timescale of order the expansion rate of the dominant vacuum, $H_{\rm max}$.  Therefore, events occurring later than a few units of $H_{\rm max}$ after the decay of the dominant vacuum have negligible probability.  This includes our own observations, so the measure is ruled out.}
\label{fig-shrink}
\end{figure} 
 The proper time measure famously suffers from the youngness problem~\cite{LinLin96,Gut00a,Gut00b,Gut04,Teg05,Lin07,Gut07}, or ``Boltzmann babies''~\cite{BouFre07}.  Typical observers are predicted to be thermal fluctuations in the early universe, and our own observations have probability of order $\exp(-10^{60})$.  This holds in any underlying landscape model as long as it contains our vacuum.  Thus the proper time measure is ruled out by observation at very high confidence level.   

Explaining the origin of the youngness problem is somewhat convoluted in the global picture.  Consider an event that occurs at 13.7 Gyr after the formation of the bubble universe it is contained in and that is included under the cutoff.  For every such event, there will be a double-exponentially large number $\exp(3H_{\rm max}\Delta\tau)$  of events in the same kind of bubble universe that occur at 13.7 Gyr$\,-\Delta\tau$ after the formation of the bubble.  This is because new bubbles of this type are produced at an exponential growth rate with characteristic time scale $H_{\rm max}$.  We will now show that the proper time cutoff has a local dual, the Incredible Shrinking Geodesic, in which the youngness problem is immediately apparent.  
 
We now seek a local dual, i.e., a geodesic with fatness (or local weight) $X(\tau)$, which will reproduce the same relative probabilities if initial conditions are chosen in the dominant (i.e., fastest-expanding) vacuum.  To find the correct fatness, we invert Eq.~(\ref{eq-tx}):
\begin{equation}
X(\tau)=e^{-3(\tau-\eta)}=\exp \int_0^\tau d\tau' \left[\theta(\tau')-3\right]~.
\label{eq-xprop}
\end{equation}
Note that this result does not satisfy the constraint we imposed in all previous subsections, that the geodesic has constant fatness in asymptotic de~Sitter regions.  
 
Obtaining a local dual in this manner is somewhat brute-force.  Recall that the duality relies on fact that the overcounting of events by overlapping fat geodesics depends only on the global time.  Here this is accomplished in two steps.  The factor $e^{3\eta}$ undoes the dilution of geodesics: it fattens the geodesics by their inverse density, thus making the overcounting factor everywhere equal to one.  The factor $e^{-3\tau}$ is a regulator that depends only on the global time and renders the integral in Eq.~(\ref{eq-nxg}) finite.  

However, the result for $X(\tau)$ immediately makes the youngness problem apparent: note that $X$ is constant as long as the geodesic remains in the fastest expanding de~Sitter vacuum, where $\theta=3H_{\rm max}=3$ (see Fig.~\ref{fig-shrink}).  However, in all other regions, $H<1$, so $\theta-3<0$ and the weight of events is suppressed exponentially as a function of the time after the decay of the dominant vacuum.  In particular, in anthropically allowed regions, such as ours, the Hubble rate is very small compared to the microscopic rate $H_{\rm max}=1$.  Thus, events are approximately suppressed as $e^{-3\tau}$, that is, exponentially with a microscopic characteristic timescale.  For example, with $H_{\rm max}$ of order the Planck scale, we thus find that events today are less likely than events yesterday by a factor of $\exp(-10^{48})$, and less likely than a billion years ago by a factor of $\exp(-10^{60})$.  As a consequence, this measure assigns higher probability to (conventionally unlikely) observers arising from large quantum fluctuations in the early universe (and their bizarre observations) than to our observations~\cite{LinLin96,Gut00a,Gut00b,Gut04,Teg05,Lin07,Gut07,BouFre07}.

\acknowledgments We would like to thank Felipe Gonzalez for help with some figures.  This work was supported by the Berkeley Center for Theoretical Physics, by the National Science Foundation (award numbers 0855653 and 0756174), by fqxi grant RFP3-1004, by a Becas-Chile scholarship, and by the U.S.\ Department of Energy under Contract DE-AC02-05CH11231.

\bibliographystyle{utcaps}
\bibliography{all}

\end{document}